\documentclass[prb,twocolumn,showpacs]{revtex4}
\usepackage{color}
\usepackage{bbm}
\usepackage{units}
\usepackage{graphicx}
\usepackage{natbib}

\begin{document}

\title{Indications for sharp continuous phase transitions at finite 
temperatures connected with the apparent metal-insulator transition in 
two-dimensional disordered systems} 

\author{Arnulf M\"obius}
\email{a.moebius@ifw-dresden.de}
\affiliation{IFW Dresden, PF 270 116, D-01171 Dresden, Germany}

\begin{abstract}
In a recent experiment, Lai {\it et al.}\ [Phys.\ Rev.\ B {\bf 75},
033314 (2007)] studied the apparent 
metal-insulator transition (MIT) of a Si quantum well structure tuning 
the charge carrier concentration $n$. They observed linear temperature 
dependences of the conductivity $\sigma(T,n)$ around the Fermi 
temperature and found that the corresponding 
$T \rightarrow 0$ extrapolation $\sigma_0(n)$ exhibits a sharp bend just 
at the MIT. Here, reconsidering the data published by Lai {\it et al.}, 
it is shown that this sharp bend is related to a peculiarity of 
$\sigma(T = {\rm const.},n)$ clearly detectable in the whole $T$ range 
up to $4\ {\rm K}$, the highest measuring temperature in that work.
Since this peculiarity seems not to be smoothed out with increasing $T$ 
it may indicate a sharp continuous phase transition between the regions 
of apparent metallic and activated conduction to be present at finite 
temperature. Hints from the literature of such a behavior are discussed. 
Finally, a scaling analysis illuminates similarities to previous 
experiments and provides understanding of the shape of the peculiarity 
and of sharp peaks found in 
$\mbox{d}\;\!\mbox{log}_{10}\;\!\sigma / \mbox{d} n\;\!(n)$. 

\end{abstract}
\date{\today}
\maketitle

\section{Introduction}

The question whether electronic conduction in two-dimensional
disordered systems is exclusively nonmetallic, so that the resistivity
$\rho(T)$ always diverges as temperature $T \rightarrow 0$, or whether 
it can also have metallic character has been under controversial debate 
for three decades. The existence of a corresponding metal-insulator 
transition (MIT) was denied by the localization theory of Abrahams {\it 
et al.},\cite{Abra.etal.79} which however neglects electron-electron 
interaction. Thus it came as a surprise when Kravchenko {\it et al.}, 
who had varied the charge carrier concentration $n$ in high mobility 
MOSFET samples, first reported a strong decrease of $\rho$ down to 
$20\ {\rm mK}$.\cite{Krav.etal.94} They considered the conduction in the 
respective $(T,n)$ region as metallic, an interpretation which was 
questioned in particular by Altshuler and Maslov,\cite{Alts.Masl.99} 
compare also Refs.\ \onlinecite{Krav.etal.99} and 
\onlinecite{Alts.Masl.99a}. The nature of the apparent metallic 
state has remained puzzling up to now, for recent reviews see Refs.\ 
\onlinecite{Krav.Sara.04,Puda.04,Sarm.Hwan.05,Shas.05}.

Provided the apparent MIT is a real phase transition at $T = 0$, it can 
be considered as a quantum critical point.\cite{Punn.Fink.05} Commonly, 
the transition is expected to be smoothed out at finite $T$. However, 
this smoothing is not a must for a quantum phase 
transition.\cite{Sach.05} For example, the Ising ferromagnet LiHoF$_4$ 
in transverse magnetic field undergoes a quantum phase transition at 
$T = 0$, where the quantum critical point is the beginning of a line of 
continuous phase transitions at finite $T$.\cite{Bitk.etal.96} Thus both 
the situations sketched in Fig.\ 1 are conceivable. It is the aim of 
this work to uncover indications in the literature for the MIT in 
two-dimensional systems being likewise connected with a line of 
continuous phase transitions at finite $T$. 

\begin{figure}
\includegraphics[width=0.71\linewidth]{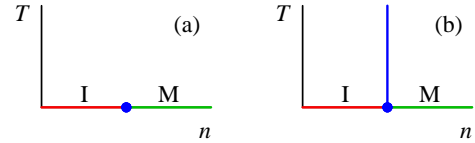}
\caption{(Color online) Two possible features of the phase diagram of a 
two-dimensional charge carrier system in the vicinity of the MIT: (a) At 
$T = 0$, a sample is either insulating (I) or metallic (M), the regions 
are separated by a quantum critical point (QCP) marked as $\bullet$. 
However, there is a smooth transition between both regimes at small, 
finite $T$. (b) The QCP is connected with a line of continuous phase 
transitions so that the MIT should be directly detectable at finite $T$ 
when $n$ is varied.}
\end{figure}

The present study is motivated by the scaling of the $T$ dependences of 
$\rho$ for various $n$, $\rho(T,n) = \rho(T/T_0(n))$, reported by 
Kravchenko {\it et al.}\ for MOSFET's.\cite{Krav.etal.95} For
$n < n_{\rm c}$, this relation is of the same type as the scaling laws 
for the hopping regions close to the metal-superconductor transition in 
ultrathin amorphous Bi films\cite{Liu.etal.91,Liu.etal.93} and close to 
the MIT in several disordered three-dimensional 
systems.\cite{Moe.etal.83,Zabr.Zino.84,Moe.etal.85,Moe.85,Sara.Dai.02}
Compare also Refs.\ \onlinecite{Liu.etal.92} and 
\onlinecite{Lam.etal.97} for ultrathin amorphous Pd films and 
for\linebreak Si-Si$_{0.87}$Ge$_{0.13}$-Si quantum-well structures, 
respectively.  

Such a scaling of the $T$ dependences of $\rho$ close to a phase 
transition, where the characteristic hopping temperature vanishes, 
implies the existence of a sharp boundary between apparent metallic and 
nonmetallic conduction concerning $\rho$, independent of $T$, and 
hence also with respect to $n$, independent of $T$. Of course, this 
statement is based on some idealizations and presumes $T$ not to exceed 
a certain threshold where additional mechanisms become important. But it 
holds also if scaling of the $T$ dependence is valid only in the region 
of activated conduction. It will be shown below that the validity of 
this type of scaling  for finite $T$ up to the transition point would 
almost always cause a peculiarity in $\rho(T = {\rm const.},n)$.

Recently, Lai and coworkers studied the transport properties close
to the apparent MIT of an n-type Si quantum well confined in a 
Si$_{0.75}$Ge$_{0.25}$/Si/Si$_{0.75}$Ge$_{0.25}$ 
heterostructure.\cite{Lai.etal.07} The authors observed the conductivity 
$\sigma$ to have nearly linear temperature dependences around the Fermi 
temperature $T_{\rm F}$ varying between roughly 2 and $2.5\ {\rm K}$, 
see Fig.\ 1 of that work. They showed that $\sigma_0(n)$, the $T=0$ 
conductivity obtained by linear extrapolation from this $T$ region, 
exhibits two regimes of different slope. Particularly important, there
is a sharp bend at the transition between the two regimes. According to 
Fig.\ 2 of Ref.\ \onlinecite{Lai.etal.07}, this sharp bend could also be
described as a knee. It coincides with the $n$ value where 
$\mbox{d} \sigma / \mbox{d} T$ changes its sign as $T \rightarrow 0$. 
The authors interpret this finding as an indication of the existence of 
two different phases.

Lai {\it et al.}\ stress the slope of $\sigma(T)$ for $T \sim T_{\rm F}$
to be almost constant within the $n$ region close to the MIT. This 
statement provokes two interesting questions: (i) Provided the 
extrapolated $\sigma_0(n)$ has a knee indeed, and the slope used in 
extrapolation is roughly constant, should not 
$\sigma(T = {\rm const.},n)$ exhibit a knee also for finite 
$T$?\linebreak (ii) If yes, is the existence of this peculiarity 
restricted to the region of linear $\sigma(T, n = {\rm const.})$ around 
$T_{\rm F}$, or is it a more general phenomenon?

A first answer is already contained in Ref.\ \onlinecite{Lai.etal.07}: 
In performing a percolation analysis, Lai and coworkers considered
$\sigma(T=0.3\ {\rm K},n)$ as best available approximation of 
$\sigma(T=0,n)$. Fig.\ 3 of their paper shows that deviations from the 
best percolation fitting occur just at the $n$ value where $\sigma_0(n)$ 
has its sharp bend. The authors considered this to support ``that the 
electronic phase in the low-density regime is different from that in the 
high-density regime''. However, Lai {\it et al.}\  did not further 
follow up this line of thought but emphasized the necessity of 
measurements at still lower temperatures.

The detailed consideration of the above two questions is the aim of the 
present work, being organized as follows. In Sec.\ II, reconsidering
data from Ref.\ \onlinecite{Lai.etal.07}, it is shown that 
several phenomenological aspects of $\sigma(T,n)$ seem to indicate 
the existence of phase transitions between apparent metallic and
activated conduction at finite temperatures. A comparison with two 
previous experiments is given in Sec.\ III. Sec.\ IV is devoted to a 
scaling analysis of the data from Ref.\ \onlinecite{Lai.etal.07}.
It yields an explanation of the shape of $\sigma(T = {\rm const.},n)$ 
close to the apparent MIT. Finally, Sec.\ V summarizes and draws 
conclusions concerning the design of future experiments.

\section{Phenomenological reconsideration of the experiment by Lai and
coworkers}

To find out to which extent and in which form the sharp bend of 
$\sigma_0(n)$ persists when $T$ is finite, I digitized Fig.\ 1(b) of 
Ref.\ \onlinecite{Lai.etal.07} by means of WinDig\cite{WinDig}, 
considering the high-resolution version of this figure available from 
the cond-mat server. Values of $\sigma(T = {\rm const.},n)$ for five 
different temperatures, 0.35, 0.64, 1.18, 2.18, and $4.0\ {\rm K}$, were 
read out in this way. They are given in Fig.\ 2 in a $\sigma$ versus $n$ 
representation. First of all, this graph shows that all 
$\sigma(T = {\rm const.},n)$ curves share a peculiar feature, namely an 
``indentation'' at $n \approx 0.32 \cdot 10^{11}\ {\rm cm}^{-2}$. It
occurs close to the concentration value where
$\mbox{d} \sigma / \mbox{d} T$ changes its sign at the lowest $T$
considered in Ref.\ \onlinecite{Lai.etal.07}, see Fig.\ 1(b) therein. 
This value, $n_{\rm c} = 0.322 \cdot 10^{11}\ {\rm cm}^{-2}$, is the 
critical concentration of the apparent MIT. It is marked by arrows in 
Figs.\ 2, 3, and 6 to 9

\begin{figure}
\includegraphics[width=0.71\linewidth]{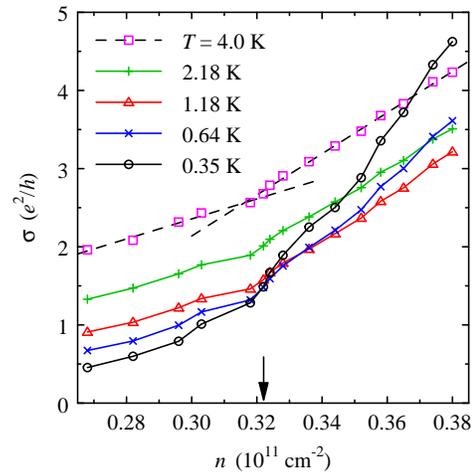}
\caption{(Color online) Charge carrier concentration dependence of the 
conductivity for the Si quantum well structure studied in Ref.\ 
\protect{\onlinecite{Lai.etal.07}}. Data were obtained from Fig.\ 1(b) 
of that work.  The arrow marks the critical concentration $n_{\rm c}$ of 
the apparent MIT, see text.}
\end{figure}

To answer the question whether or not the sharp bend observed by Lai 
{\it et al.}\ at $\sigma_0(n)$ persists at finite $T$, fits by two 
linear functions, concerning the $n$ regions below and above 
$0.32 \cdot 10^{11}\ {\rm cm}^{-2}$, were tried. Fig.\ 2 includes such 
linear fits for the case $T = 4.0\ {\rm K}$. Note, how remarkably well 
the two dashed lines describe the measured data. 

However, there is a small inconsistency in this approximation: The two 
linear functions intersect at $n = 0.317 \cdot 10^{11}\ {\rm cm}^{-2}$, 
a slightly lower value than the interval splitting concentration. This 
problem is avoided when approximating the experimental data points by 
the piecewise linear function 
$f_{\rm pl}(n) = a + b n + c (n - n_{\rm k}) \theta (n - n_{\rm k})$ 
with the adjustable parameters $a$, $b$, $c$, and $n_{\rm k}$. Here, 
$n_{\rm k}$ denotes the knee position, and $\theta$ stands for the 
Heaviside step function. When included in Fig.\ 2, such a piecewise 
linear fit of $\sigma(T = 4.0\ {\rm K},n)$ would deviate from the dashed 
lines by less than the line width. So it is not shown for clarity 
reasons.

For the piecewise linear fit of $\sigma(T = 4.0\ {\rm K},n)$, the least 
square sum $\chi^2$ of the deviations between fit and data points 
amounts to $0.010\ e^4/h^2$ when weight 1 is ascribed to all points. For 
alternative approximations by polynomials of second or third order, one 
obtains clearly worse values, $\chi^2 \approx 0.029\ e^4/h^2$ and 
$0.024\ e^4/h^2$, respectively. This is a first indication for the MIT 
being connected with a sharp continuous phase transition at finite $T$. 

Also the curves for the other $T$ values can be nicely approximated by 
piecewise linear functions. Tab.\ I presents the optimum knee positions 
$n_{\rm k}$ and the related $\chi^2$ values. It is remarkable that the 
same $n_{\rm k}$ value, $0.318 \cdot 10^{11}\ {\rm cm}^{-2}$, results for
all temperatures but $T = 0.35\ {\rm K}$. This value is only slightly 
lower than the abscissa of the intersection for the $T \rightarrow 0$ 
extrapolation $\sigma_0(n)$ in Fig.\ 2 of Ref.\ 
\onlinecite{Lai.etal.07}. 

However, when comparing with polynomial approximations for the total $n$ 
region, it is noted that the advantage of the approximation by piecewise 
linear functions over that by one polynomial of third order (same number 
of adjustable parameters) diminishes with decreasing $T$: The 
approximation by piecewise linear functions is clearly advantageous for 
4.0, 2.18, and $1.18\ {\rm K}$, it has roughly the same quality as 
the polynomial approximation for $T = 0.64\ {\rm K}$, and it is clearly
worse than the polynomial approximation for $T = 0.35\ {\rm K}$, see 
Table I. This trend is not surprising because 
$\sigma(T = {\rm const.},n)$ is basically a nonlinear function in the 
region of activated conduction, it vanishes in some exponential way with 
decreasing $n$. Hence, when considering a fixed $n$ range and lowering 
$T$, the $\sigma$ region (on a logarithmic scale) broadens, and the 
basic nonlinearity gets increasing weight compared to the 
``indentation'' at $n \approx 0.32 \cdot 10^{11}\ {\rm cm}^{-2}$. 

\begin{table}
\caption{Results of phenomenological fits of 
$\sigma(T = {\rm const.},n)$ by piecewise linear functions (plf) and 
polynomials of third order (pto): Values of knee positions $n_{\rm k}$ 
and least square sums $\chi^2$ are given in units of 
$10^{11}\ {\rm cm}^{-2}$ and $e^4/h^2$, respectively. The left group of 
columns results from consideration of the complete data sets, the right 
group is obtained from the data points fulfilling
$0.5\,\sigma(T,n_{\rm c}) < \sigma(T,n) < 2.0\,\sigma(T,n_{\rm c})$. 
$N$ denotes the number of data points taken into account.}
\begin{ruledtabular}
\begin{tabular}{c|cccc|cccc}
 & \multicolumn{4}{c|}{complete data set} 
 & \multicolumn{4}{c}{restricted data set}\\ \hline
 $T({\rm K})$ & $N$ & $n_{\rm k}$ & $\chi^2_{\rm plf}$ & 
 $\chi^2_{\rm pto}$ & $N$ & $n_{\rm k}$ & $\chi^2_{\rm plf}$ & 
 $\chi^2_{\rm pto}$\\
\hline
 4.0 & 15 & 0.318 & 0.010 & 0.024 & 15 & 0.318 & 0.010 & 0.024 \\
 2.18 & 15 & 0.318 & 0.006 & 0.019 & 15 & 0.318 & 0.006 & 0.019 \\
 1.18 & 15 & 0.318 & 0.010 & 0.019 & 14 & 0.318 & 0.009 & 0.018 \\
 0.64 & 15 & 0.318 & 0.028 & 0.031 & 11 & 0.318 & 0.013 & 0.026 \\
 0.35 & 15 & 0.315 & 0.097 & 0.063 & 9 & 0.318 & 0.024 & 0.036 \\
\end{tabular}
\end{ruledtabular}
\end{table}

A more meaningful comparison of both approximations is obtained 
restricting all fits to comparable $\sigma$ intervals, namely to 
$\{0.5\,\sigma(T,n_{\rm c}),2.0\,\sigma(T,n_{\rm c})\}$.  Under this 
condition, for each of the $T$ values considered in Fig.\ 2, piecewise 
linear functions clearly approximate the data better than a polynomial 
of third order, see Table I. Moreover, now all fits yield the same 
$n_{\rm k}$ value, $0.318 \cdot 10^{11}\ {\rm cm}^{-2}$.

Alternative fits concerning the conductivity region 
$\{0.5\,\sigma(T,n_{\rm k0}), 2.0\,\sigma(T,n_{\rm k0})\}$
with $n_{\rm k0} = 0.318 \cdot 10^{11}\ {\rm cm}^{-2}$ yield very 
similar results: The advantage of the piecewise linear fits is confirmed 
for all $T$. The only change of the $n_{\rm k}$ value occurs for 
$0.64\ {\rm K}$ where the alternative fit results in
$0.316  \cdot 10^{11}\ {\rm cm}^{-2}$.

Nevertheless, one should not overestimate the piecewise linear fits. The 
conclusive message from their advantage over the polynomial fits is only 
that, for all $T$ values, some characteristic change occurs within a 
very small $n$ region close to the knee of the piecewise linear fit. In 
this sense, the sharp bend observed by Lai {\it et al.}\ at the 
$T \rightarrow 0$ extrapolation $\sigma_0(n)$ turns out to be not 
smoothed for finite $T$. Moreover, it seems to exist even for 
$T \ll T_{\rm F}$, where $\sigma(T,n = {\rm const.})$ is nonlinear and 
considerably deviates from the estimate by means of the linear 
extrapolation from the vicinity of $T_{\rm F}$.

It is enlightening to compare with the behavior of 
$\mbox{d} \sigma / \mbox{d} T$ as $T \rightarrow 0$ according to Fig.\ 
1(b) of Ref.\ \onlinecite{Lai.etal.07}. The sign change of 
$\mbox{d} \sigma / \mbox{d} T$ at 
$n_{\rm c} = 0.322 \cdot 10^{11}\ {\rm cm}^{-2}$ and 
$\sigma_{\rm c} \approx 1.5\;\!h/e^2$, defining the apparent MIT, 
roughly coincides with the sharp bends in $\sigma(T = {\rm const.},n)$. 
Thus it is likely that the bends originate from sharp continuous phase 
transitions separating the regions of apparent metallic and activated 
conduction at finite $T$. The related critical concentration seems to be 
nearly independent of $T$. 

The values of $n_{\rm k}$ and $n_{\rm c}$ slightly differ from each 
other.  The small discrepancy may arise from the approximation of 
$\sigma(T = {\rm const.},n)$ by a piecewise linear function being an 
oversimplification. This is also suggested by inspecting the data points 
in the immediate vicinity of the apparent MIT in Fig.\ 2. 

Thus a more detailed investigation is desirable. An alternative 
presentation of the ``indentation'' with a better resolution is given 
for that purpose in Fig.\ 3. Here, the resistivity $\rho$ in dependence 
on $n$ is shown (so that the ``indentation'' is directed ``upwards''). A 
$\mbox{log}_{10}\;\!\rho$ scale is used as in many experimental papers 
on the apparent MIT in two-dimensional systems. However, this graph 
focuses on a rather small concentration region, $n$ merely varies 
between 0.296 and $0.352 \cdot 10^{11}\ {\rm cm}^{-2}$. For optimum 
resolution, each $\rho(T)$ curve is shown in a separate subgraph where 
its average slope is scaled to a common value. In order to make details 
of the curves better visible, two purely phenomenological fits are 
included: a linear regression (dashed) based on the 4 points from 0.328 
to $0.352 \cdot 10^{11}\ {\rm cm}^{-2}$ and a regression parabola 
(dash-dotted) for the 6 points from 0.318 to 
$0.344 \cdot 10^{11}\ {\rm cm}^{-2}$. In these fits, 
$\mbox{log}_{10}\;\!\rho$ is considered as function of $n$.

Fig.\ 3 shows that deviations from the two fits occur rather
suddenly, independently of $T$ always at the same concentration values. 
Thus $\rho(T = {\rm const.},n)$ seems to exhibit three regions of 
differing behavior: below $0.318 \cdot 10^{11}\ {\rm cm}^{-2}$, between 
0.318 and $0.328 \cdot 10^{11}\ {\rm cm}^{-2}$, and above 
$0.328 \cdot 10^{11}\ {\rm cm}^{-2}$. The medium part resembles a 
rounded step. Remarkably, its width seems to be almost independent of 
$T$. Its amplitude at $n = 0.318 \cdot 10^{11}\ {\rm cm}^{-2}$ can be
estimated by means of comparing with the linear extrapolation (of 
$\mbox{log}_{10}\;\!\rho$ in dependence on $n$) from the $n$ region 
above $0.328 \cdot 10^{11}\ {\rm cm}^{-2}$. It decreases from roughly 
$26\ \%$ to roughly $5\ \%$ as $T$ increases from 0.35 to 
$4.0\ {\rm K}$. 

Due to its $T$-independent width, the ``rounded step'' may indicate a 
(with respect to the experimental resolution) sharp continuous phase 
transition occurring at finite $T$. Note that the sign change of
$\mbox{d} \sigma / \mbox{d} T$ as $T \rightarrow 0$, marked by the 
arrow, occurs roughly at the middle of the ``rounded step'' in 
$\rho(T = {\rm const.},n)$. This coincidence can be regarded as evidence 
for the hypothetical phase transition being the finite temperature 
extension of the zero-temperature effect metal-insulator transition.

\begin{figure}
\includegraphics[width=0.71\linewidth]{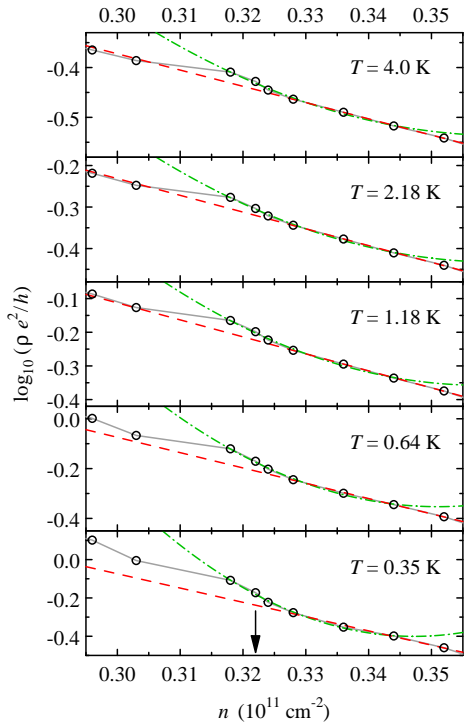}
\caption{(Color online) Details of the dependence of the resistivity on 
the charge carrier concentration in the immediate vicinity of the MIT 
for Si quantum well data shown in Fig.\ 2. For the explanation of the 
dashed and dash-dotted lines see text.}
\end{figure}

In order to exclude the possibility that the peculiarity might be 
pretended by the specific choice of the phenomenological regression 
curves included in Fig.\ 3, linear regressions for the intervals from 
0.282 to 0.303 and from 0.318 to $0.328 \cdot 10^{11}\ {\rm cm}^{-2}$ 
were tried as an alternative. From this perspective, the 
$\sigma(T = {\rm const.},n)$ data sets exhibit the same characteristics 
as in Fig.\ 3: There are three $n$ regions, clearly differing concerning 
the slope of $\mbox{log}_{10}\;\!\rho(n)$. The apparent MIT falls in the 
middle region, the width of which is roughly constant again. This 
supports the above proposed interpretation in terms of a phase 
transition between activated and apparent metallic conduction at finite 
$T$. Because of its similarity to Fig.\ 3, the corresponding graph is 
not shown here.

\section{Comparison with other experiments}

One could object the above analysis furthermore with the following 
argument: Even if the peculiarity under discussion is clearly visible at 
five different temperatures, it might be an artifact of some systematic
experimental imperfections specific to 
Ref.\ \onlinecite{Lai.etal.07}. This objection could also be formulated 
as the question whether or not similar peculiarities have been reported 
in previous experimental studies. In fact, to the best of my knowledge, 
this is not the case. 

\begin{figure}
\includegraphics[width=0.71\linewidth]{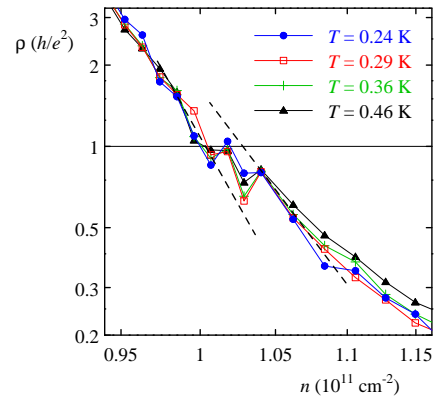}
\caption{(Color online) log-log diagram of 
$\rho(T = {\rm const.},n)$ for the intermediate vicinity of the apparent 
MIT in a high mobility MOSFET, reproduced from Fig.\ 2 of Ref.\ 
{\protect \onlinecite{Moe.02}}. Data were obtained from Fig.\ 1 of Ref.\ 
\protect{\onlinecite{Krav.etal.95}}. The dashed lines serve only as 
guide to the eye.}
\end{figure}

\begin{figure}
\includegraphics[width=0.71\linewidth]{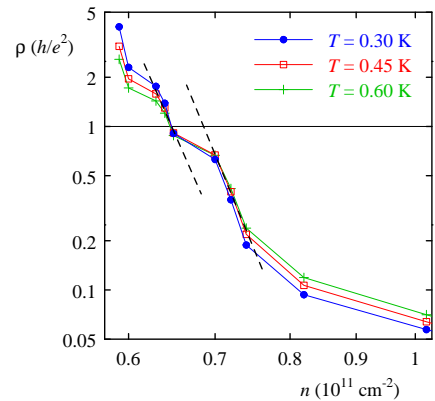}
\caption{(Color online) log-log diagram of 
$\rho(T = {\rm const.},n)$ for the vicinity of the apparent MIT in an 
AlAs quantum well, reproduced from Fig.\ 3 of Ref.\ {\protect 
\onlinecite{Moe.02}}. Data were obtained from Fig.\ 2 of Ref.\ 
\protect{\onlinecite{Papa.Shay}}, considering the high-mobility 
direction. The dashed lines are included as guide to the eye.}
\end{figure}

However, a reconsideration comparing experimental $\rho(T)$ data for the 
apparent MIT in two-dimensional systems with experience on 
three-dimensional disordered systems was performed five years 
ago.\cite{Moe.02} It points to possible peculiarities in 
$\rho(T = {\rm const.},n)$ close to the apparent MIT in measurements
at an high mobility Si MOSFET\cite{Krav.etal.95} by Kravchenko {\it et 
al}.\ and at an AlAs quantum well \cite{Papa.Shay} by Papadakis and 
Shayegan. The respective graphs from Ref.\ \onlinecite{Moe.02} are 
reproduced here as Figs.\ 4 and 5. 

Fig.\ 4 shows that already in the results of Ref.\ 
\onlinecite{Krav.etal.95}, one of the first studies of the apparent MIT, 
a behavior may be hidden which is similar to that in Fig.\ 3: The 
$\rho(T = {\rm const.}, n)$ curves may exhibit a shoulder-like structure 
close to $\rho = h/e^2$ being related to an offset of the curve parts 
below and above $h/e^2$. This structure resembles the ``rounded step'' 
in Fig.\ 3 discussed above. It is located close to the intersection 
region of the $\rho(T = {\rm const.}, n)$ curves, for a more detailed 
discussion see Ref.\ \onlinecite{Moe.02}.  Also the AlAs quantum 
well studied by  Papadakis and Shayegan\cite{Papa.Shay} seems to exhibit 
such an offset in $\rho(T = {\rm const.}, n)$ close to the apparent MIT 
as Fig.\ 5 displays. 

One point is particularly remarkable: For the lowest considered $T$, the 
highest resistivity values of the high $n$ branch of 
$\rho(T = {\rm const.},n)$ are roughly the same in the three experiments 
considered here. They amount to 0.78, 0.80, and $0.63\ h/e^2$ for Refs.\ 
\onlinecite{Lai.etal.07}, \onlinecite{Krav.etal.95}, and 
\onlinecite{Papa.Shay}, respectively.

Concluding the comparison, Figs.\ 4 and 5 resemble Fig.\ 3 
substantially, but they are certainly not as conclusive as the 
reconsideration of data from Ref.\ \onlinecite{Lai.etal.07} above. 
Nevertheless, in my opinion, they give additional support to the 
hypothesis of sharp continuous phase transitions at finite 
temperatures.

For all the three experiments reconsidered here, one could of course 
object that the emphasized features might arise from random deviations 
in the measured values of $n$ and $\rho$. However, it seems to be very 
unlikely that all the following coincidences occur only by chance:

(i) These features seem to be present for all temperatures considered.

(ii) They are observed at data from three independent experiments with
different kinds of samples, made up of different materials.

(iii) The peculiarities have qualitatively the same form in all three 
data sets.

(iv) They occur in all cases in the same resistivity region, slightly 
below $h/e^2$.

(v) They are observed in all cases close to the common intersection
point of the $\rho(T = {\rm const.},n)$ curves.

It has to be mentioned that similar hints to a possible peculiarity
could not be found in data from other publications, for example Ref.\ 
\onlinecite{Puda.etal.01}. However, this is not a strong objection to 
the reconsideration presented here for the following reasons. 

Possibly, on the one hand, very low charge carrier densities together 
with extremely high mobilities as realized in Ref.\ 
\onlinecite{Lai.etal.07} might be needed. On the other hand, the 
peculiarities may be easily overlooked: Many of the experimental studies 
in the literature concern broad $n$ ranges, and do not obtain so many 
data points in the immediate vicinity of the apparent MIT, at both 
sides, as Ref.\ \onlinecite{Lai.etal.07}. Thus the peculiarity is small 
on the logarithmic $\rho$ scale usually used and may easily be covered 
by the nonlinear $n$ dependence of $\sigma$, compare the effect of 
restricting the $\sigma$ range discussed in Sec.\ II. Moreover, when 
considering $\rho$ on a logarithmic scale, there is no need to achieve a 
very high precision of the individual data points, so that the small 
peculiarity may remain hidden behind random deviations. Additionally, 
arbitrary sample inhomogeneities including edge effects tend to smear 
it, see the careful study in Ref.\ \onlinecite{Lilly.etal.02}, in 
particular Fig.\ 4. The last three reasons seem to me more likely 
explanations than the possible necessity of an extremely high mobility.

\section{Scaling analysis}

Provided the ``rounded step'' is indeed related to a phase transition, 
how may it arise? On a macroscopic level, it can be explained in 
the following way: Suppose, on the insulating side of the apparent MIT, 
the $T$ dependence of $\sigma$ scales as observed by Kravchenko 
{\it et al.}\ at the resistivity of MOSFET's:\cite{Krav.etal.95}
\begin{equation}
\sigma(T,n) = \sigma(t)\ \ {\rm with}\ \ t = T / T_0(n)\,,
\end{equation}
for further references concerning this matter see the introduction. To 
the best of my knowledge, there is no systematic study how this scaling 
is softened with increasing $T$ close to the MIT. So, according to the 
above hypothesis, assume that it holds up to the critical concentration 
$n_{\rm c}$. Since $\mbox{d} \sigma / \mbox{d} T \rightarrow 0$ as
$n \rightarrow n_{\rm c}$, $T_0(n)$ vanishes and  $t$ diverges as 
$n \rightarrow n_{\rm c}$, independent of $T$. This is connected with 
$\sigma(T,n_{\rm c}) = \sigma(t = \infty) = \sigma_{\rm c}$ (provided 
$T$ does not exceed a certain threshold where other mechanisms become 
relevant). 

Suppose furthermore
\begin{equation}
T_0(n) = A |\delta n|^{\beta} \ \ {\rm with}\ \ 
\delta n = n - n_{\rm c}\,,
\end{equation}
where $A$ is a constant which might, however, be sample dependent. The 
value of the critical exponent was determined by Kravchenko {\it et 
al.}\ by means of a scaling analysis of their MOSFET experiment 
yielding $\beta = 1.6 \pm 0.1$.\cite{Krav.etal.95} This result was 
confirmed by Lam {\it et al.}\ for a Si-Si$_{0.87}$Ge$_{0.13}$-Si 
quantum well, $\beta = 1.6 \pm 0.2$.\cite{Lam.etal.97} (The comparison 
with the behavior of ultrathin metal films analyzed in Ref.\ 
\onlinecite{Liu.etal.93} is hindered by the use of the resistance as 
control parameter in that work.)

According to Eqs.\ (1) and (2), $\sigma$ should depend only on the 
parameter $T/| \delta n |^{\beta}$ or rather on $\delta n/T^{1/\beta}$. 
This can be checked analogously to e.g.\ Fig.\ 10 of Ref.\ 
\onlinecite{Krav.etal.95} or Fig.\ 3 of Ref.\ \onlinecite{Pare.etal.05}.
Since the validity of Eq.\ (2) is needed, such a scaling check is less
general than the construction of a mastercurve by rescaling of $T$. 
However, it has a big advantage: Natural small sample inhomogeneities 
$\Delta n$ do not destroy a possible generic coincidence of the curves 
for $|\delta n| \gg \Delta n$, although such inhomogeneities might even
cause nonmonotonic behavior of $\sigma(T,n = {\rm const.})$ in the 
immediate vicinity of $n_{\rm c}$.

A check of the hypothesis that $\sigma$ is a function of 
$\delta n/T^{1/\beta}$ alone is given in Fig.\ 6 considering the 
$\sigma$ curves for $T = 0.35$ and $0.64\ {\rm K}$ from Fig.\ 2. For 
that plot, $n_{\rm c} = 0.322 \cdot 10^{11}\ {\rm cm}^{-2}$ 
corresponding to the sign change of $\mbox{d} \sigma / \mbox{d} T$ as 
$T \rightarrow 0$, and $\beta = 1.6$ as obtained in Ref.\ 
\onlinecite{Krav.etal.95} are utilized. Curves for higher $T$ are not 
taken into account since $\sigma(T,n_{\rm c})$ becomes $T$ dependent 
above roughly $1\ {\rm K}$, presumably because some additional mechanism 
becomes relevant there, see Fig.\ 1 of Ref.\ \onlinecite{Lai.etal.07}.

Fig.\ 6 shows that both $\sigma$ curves nicely fall together. Note also 
the conformity concerning the ``indentation'' around $\delta n = 0$. 
Since the transformation used here does not include any adjustable 
parameter, it is unlikely that the agreement of both curves arises only 
by chance. Thus Fig.\ 6 supports the scaling equations (1) and (2).

\begin{figure}
\includegraphics[width=0.71\linewidth]{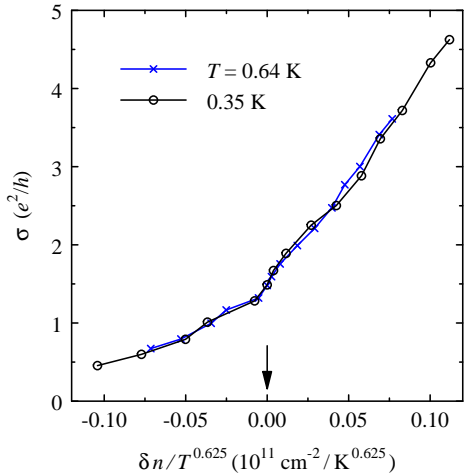}
\caption{(Color online) Check regarding scaling behavior according to 
Eqs.\ (1) and (2) for the two $\sigma$ curves from Fig.\ 2, related to 
the lowest $T$ values, 0.35 and $0.64\ {\rm K}$. Note: The agreement of 
both curves does not result from adjusting parameters.}
\end{figure}

\begin{figure}
\includegraphics[width=0.71\linewidth]{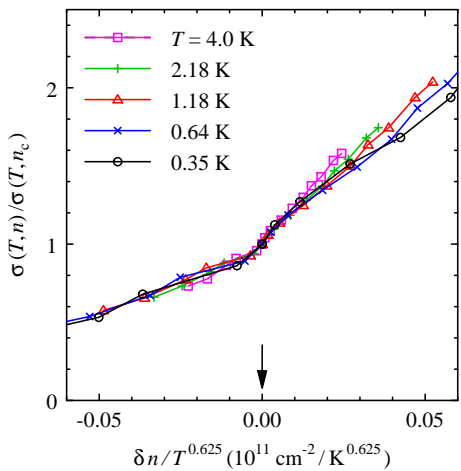}
\caption{(Color online) Scaling check with respect to Eq.\ (2) and (4) 
for all $\sigma$ curves presented in Fig.\ 2, presuming $\beta = 1.6$.}
\end{figure}

Now the question is suggested whether signs of the scaling demonstrated 
in Fig.\ 6 can be detected also for $T > 1\ {\rm K}$ where 
$\sigma(T,n_{\rm c})$ has a significant $T$ dependence. Here, a 
comparison with the MIT in three-dimensional systems is helpful. It was
observed at amorphous Si$_{1-x}$Cr$_x$ films\cite{Moe.etal.85} that,
close to $n_{\rm c}$, the multiplicative decomposition  
\begin{equation}
\sigma(T,n) = \sigma_{\rm scal}(T/T_0(n)) \cdot \xi(T,n)
\end{equation}
is useful, where $\sigma_{\rm scal}$ denotes the scaling contribution 
valid in the low-temperature region, and $\xi$, with $\xi \rightarrow 1$ 
as $T \rightarrow 0$, describes the high-$T$ deviations from the scaling 
behavior. Fig.\ 5 of Ref.\ \onlinecite{Moe.etal.85} shows that 
$\xi(T,n)$ proved to be almost independent of $n$ in the vicinity of 
the MIT. Presuming for the moment that $\xi$ is indeed independent of 
$n$, Eq.\ (3) implies
\begin{equation}
\sigma(T,n) / \sigma(T,n_{\rm c}) = 
\sigma_{\rm scal}(T/T_0(n)) / \sigma_{\rm c}\,.
\end{equation}
In this sense, for the data considered here, the ratio 
$\sigma(T,n) / \sigma(T,n_{\rm c})$ might be a function of 
$\delta n/T^{1/\beta}$ alone within the whole $T$ region from 0.35 to
$4.0\ {\rm K}$. This hypothesis is checked in Figs.\ 7 and 8 assuming 
$n_{\rm c} = 0.322 \cdot 10^{11}\ {\rm cm}^{-2}$ as in Fig.\ 6. 

\begin{figure}
\includegraphics[width=0.71\linewidth]{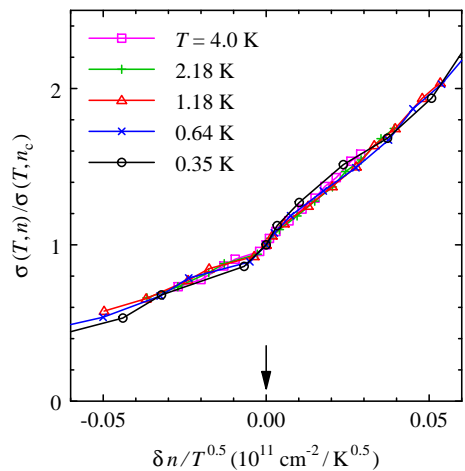}
\caption{(Color online) Scaling check with respect to Eqs.\ (2) and (4) 
for all $\sigma$ curves presented in Fig.\ 2, presuming $\beta = 2.0$.}
\end{figure}

Fig.\ 7 is based on $\beta = 1.6$ as Fig.\ 6. It shows indeed an 
approximate collapse of the data for various $T$ on one curve: Except 
for $T = 4.0\ {\rm K}$, all points with 
$\delta n / T^{0.625} < 0.03 \cdot 10^{11} \ {\rm cm}^{-2} 
{\rm K}^{-0.625}$ can be considered to fall onto one curve. However, for
larger values of $\delta n / T^{0.625}$, there are clear deviations. 
They can have several causes: (a) The scaling equation (1) is expected 
to be violated somewhere in the region of metallic conduction. 
Otherwise, $\sigma(T=0,n)$ would have to be independent of $n$. (b) The 
deviations may originate from a weak $n$ dependence of $\xi$. (c) 
Inaccuracies of the values of $n_{\rm c}$ and $\beta$ may be the reason.

Fig.\ 8 demonstrates that a shift of $\beta$ from 1.6 to 2.0 leads to a
far better data collapse. (Its quality is once more considerably 
improved if the points for the lowest temperature, $T = 0.35\ {\rm K}$, 
are omitted.) However, it remains an open question at present whether 
the improved agreement of the curves has a physical meaning or arises 
from the compensation of several deviations from scaling.

Nevertheless, in collapsing points for various $T$ values onto one 
curve, Figs.\ 6 to 8 clearly testify that, independently of $T$, all 
$\sigma(T = {\rm const.},n)$ curves exhibit a sharp bend very close to
$n_{\rm c}$. Thus they support once more the hypothesis of a line of 
phase transitions at finite $T$. Simultaneously, the above consideration
has the consequence that $\sigma(T,n_{\rm c})$ has the meaning of a 
minimum metallic conductivity also for $T > 1\ {\rm K}$ even though it 
is $T$ dependent in this region.

Although the argumentation in this chapter is solely based on scaling 
according to Eq.\ (1) within the region of activated conduction, the 
behavior on the apparent metallic side as shown in Figs.\ 6 to 8 
likewise deserves attention. These graphs, in particular Fig.\ 6, may 
indicate analogous scaling behavior also for a certain $n$ region above 
$n_{\rm c}$, in agreement with the original findings in Ref.\ 
\onlinecite{Krav.etal.95}, but in contradiction to later publications. 
Since such a scaling would not be consistent with usual metallic 
transport, systematic experimental studies of how this scaling breaks 
down should be very interesting.

The quantitative determination of the critical exponent $\beta$ is 
hampered by the lack of a theoretically based analytic expression for 
$\sigma_{\rm scal}(t)$ being applicable to a broad $\sigma$ region. The 
only resort is to use a sufficiently precise empirical description of 
the hypothetical master curve $f(\delta n/T^{1/\beta}) = r(T,\delta n) = 
\sigma(T,n)/\sigma(T,n_{\rm c})$ with an as small as possible number of 
adjustable parameters. In practice, for several temperature pairs
$(T_1,T_2)$, the value of the scaling factor $X = (T_1/T_2)^{1/\beta}$ 
is obtained by ``moving'' the experimental curve  $r(T_2,X \delta n)$ on 
top of the curve $r(T_1,\delta n)$. This is realized approximating 
$r(T_1,\delta n)$ and $r(T_2,X \delta n)$ by a common empirical 
$g(\delta n)$ with several adjustable parameters. In minimizing the 
total mean square deviation of both measured data sets from 
$g(\delta n)$, the parameters of $g$ and the scaling factor $X$ are 
adjusted.

This procedure is complicated by the sharp bend in 
$\sigma(T = {\rm const.},\delta n)$: For example, a polynomial would 
have to be of a very high degree to emulate this feature. A way out is 
to ``uncompress'' the immediate vicinity of $n_{\rm c}$ by considering 
$z = {\rm sign}(\delta n) |\delta n|^{1/2}$ as independent variable 
instead of $\delta n$. Thus, as various trials corroborated, 
${\overline r}(T,z) = r(T,\delta n)$ can be well approximated by a
${\overline g}(z)$ being a polynomial of rather low degree or a natural 
cubic spline with a rather small number of equidistant nodes.

Table II presents values of the scaling factor $X$ obtained in a series 
of such polynomial regression analyses. The errors of $X$ given are 
rough estimates of the $1 \sigma$ bound presuming that the deviations 
from ${\overline g}(z)$ are only of random nature, and that the relative 
errors have Gaussian distributions of same width for all data points 
taken into account. For four pairs $(T_1,T_2)$, the table compares the 
result of investigating the total ${\overline r}(T,z)$ data sets with 
the two values obtained when omitting the four outer points with 
$|\delta n| > 0.04  \cdot 10^{11}\ {\rm cm}^{-2}$ from each 
experimental curve, and when considering only the region of activated 
conduction, $|\delta n| \le 0$. In these three cases, polynomials of
degree 5, 4, and 3, respectively, were used as ${\overline g}(z)$. 
Additional analyses based on polynomials of by 1 to 2 enlarged degree 
yielded very similar results, as did studies based on natural cubic
splines.

\begin{table}
\caption{Results of scaling regression analyses determining the best 
overlap of the curves 
$\sigma(T_1,n_{\rm c} + \delta n) / \sigma(T_1,n_{\rm c})$ and 
$\sigma(T_2,n_{\rm c} + X\, \delta n) / \sigma(T_2,n_{\rm c})$. The
three columns of $X$ values refer to the study of complete $\sigma(T,n)$
data sets, to the analysis of data sets restricted by the demand 
$|\delta n| \le 0.04  \cdot 10^{11}\ {\rm cm}^{-2}$, and to the 
consideration of only the points within the region of activated 
conduction, $|\delta n| \le 0$.}
\begin{ruledtabular}
\begin{tabular}{c|c|c|c|c}
 $T_1({\rm K})$ & $T_2({\rm K})$ & $X(\rm compl.\ set)$ & 
 $X(\rm restr.\ set)$ & $X(\rm activ.\ cond.)$ \\
\hline
 0.35 & 0.64 & 0.684 $\pm$ 0.017 & 0.693 $\pm$ 0.027 & 0.646 $\pm$ 0.037 \\
 0.64 & 1.18 & 0.719 $\pm$ 0.015 & 0.729 $\pm$ 0.021 & 0.686 $\pm$ 0.045 \\
 1.18 & 2.18 & 0.744 $\pm$ 0.011 & 0.759 $\pm$ 0.016 & 0.770 $\pm$ 0.037 \\
 2.18 & 4.0  & 0.788 $\pm$ 0.012 & 0.811 $\pm$ 0.016 & 0.781 $\pm$ 0.038 \\
\end{tabular}
\end{ruledtabular}
\end{table}

Table II shows that, for all $(T_1,T_2)$ pairs, the three $X$ values 
agree nicely with each other. Since the $r(T,\delta n)$ curves have a 
considerable curvature and, moreover, exhibit the sharp bend fine 
structure, this agreement supports the phenomenological Eqs.\ (1) to
(4) within experimental accuracy in the following sense: The $n$ 
dependence of $\xi$ seems to be very weak indeed, otherwise the values 
in the first two $X$ columns would have to differ systematically. 
Moreover, the agreement between first and third $X$ columns supports the 
hypothesis obtained above from Figs.\ 6 to 8 that scaling behavior may 
be valid also in a certain $n$ region of apparent metallic conduction. 

The question remains whether the deviations from scaling in Fig.\ 7 may 
arise mainly from using an imprecise value of $\beta$, or whether Figs.\ 
7 and 8 point to systematic deviations from scaling indicated by a 
feigned drift of $\beta$. This exponent can be directly obtained from 
the $X$ values in Table II.  However, in estimating its error bound, one 
has to have in mind that the measurements in Ref.\ 
\onlinecite{Lai.etal.07} were performed in a dynamic modus with slowly 
varying temperature, see Fig.\ 1 therein. Additionally to calibration 
uncertainties, such a procedure exhibits some small systematic error of 
the $T$ values arising from imperfect equilibration. Even though these 
inaccuracies are so small, that they are extraneous to the data analysis 
in Ref.\ \onlinecite{Lai.etal.07}, they may have a significant influence 
on the $\beta$ values obtained here. Thus, to judge the influence of 
systematic $T$ errors, the corresponding uncertainties of $\beta$ were 
estimated, cautiously guessing $\Delta T/T \sim 0.03$. 

In this way, the $\beta$ values $1.6 \pm 0.5$, $1.9 \pm 0.5$, 
$2.1 \pm 0.5$, and $2.5 \pm 0.7$ are obtained considering the intervals
\linebreak 0.35--0.64 K, 0.64--1.18 K, 1.18--2.18 K, and 2.18--4.0 K, 
respectively. Here, the error originates from the $3 \sigma$ uncertainty 
of $X$ and from the systematic $T$ deviations.  According to these 
results, the effective $\beta$ value might slowly vary with $T$ due to 
some deviations from the very simple Eqs.\ (1--4). However, for the size 
of the error bars, this variation of $\beta$ can still not be taken for 
certain. Further experiments are needed to clarify this question.

We focus now again on the region of activated conduction. For the 
immediate vicinity of the MIT, that means for large $t$ and 
$|\sigma_{\rm c} - \sigma|/\sigma_{\rm c} \ll 1$, the following ansatz 
is suggested:
\begin{equation}
\sigma(t) = \sigma_{\rm c} \cdot (1 - B\, t^{-\nu})
\end{equation}
Here $B$ is a dimensionless constant and $\nu$ a positive exponent. 
Unfortunately, only a very rough guess of the value of $\nu$ can be 
given at present. If the value would correspond to hopping in the 
Coulomb gap for small $t$, $\nu$ would be 1/2. However, according to 
Ref.\ \onlinecite{Krav.etal.95}, $\nu$ may be expected to be smaller
for large $t$. In both cases, for $\beta = 1.6$ according to Ref.\ 
\onlinecite{Krav.etal.95}, the product $\beta \nu$ would be clearly 
smaller than 1.

Equations (1), (2), and (5) determine the type of the $n$ dependence of 
$\sigma$ close to $\sigma_{\rm c}$:
\begin{equation}
\sigma(T,n) = \sigma_{\rm c} \cdot (1 - C\, |\delta n|^{\beta \nu})
\end{equation}
where
\begin{equation}
C = B (A/T)^{\nu}\,.
\end{equation}
Thus, only if $\beta \nu = 1$, $\sigma(T = {\rm const.},n)$ does not 
exhibit any specific feature when $n$ approaches $n_{\rm c}$ from 
the side of activated conduction; nevertheless, 
$\sigma(T = {\rm const.},n)$ may have a knee at $n_{\rm c}$ in this 
special situation. In case $\beta \nu < 1$ as guessed above, 
$\sigma(T = {\rm const.},n)$ has a root-like peculiarity at $n_{\rm c}$, 
and $\rho(T = {\rm const.},n)$ as well. Taking into account small sample 
inhomogeneities, this peculiarity is softened to some ``rounded step''. 
This is in qualitative agreement with Fig.\ 3. 

\begin{figure}
\includegraphics[width=0.71\linewidth]{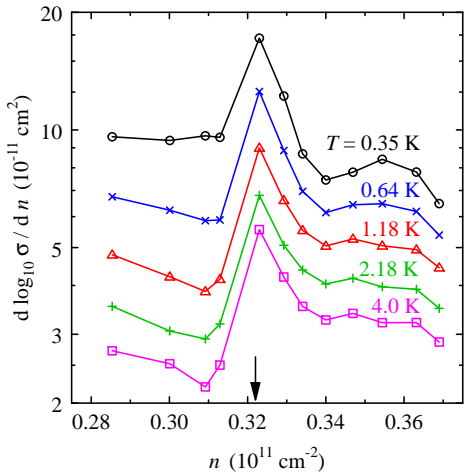}
\caption{(Color online) 
$\mbox{d}\;\!\mbox{log}_{10}\;\!\sigma / \mbox{d} n$ versus $n$ for all 
curves displayed in Fig.\ 2.}
\end{figure}

A root-like peculiarity in $\sigma(T = {\rm const.},n)$ would imply a 
divergence of $\mbox{d} \sigma / \mbox{d} n$, and thus also of 
$\mbox{d}\;\!\mbox{log}_{10}\;\!\sigma / \mbox{d} n$. (Considering 
$\mbox{log}_{10}\;\!\sigma$ rather than $\sigma$ has the advantage that 
the contribution from $\xi$, see Eq.\ (3), is suppressed.) The 
divergence would be smoothed to a sharp maximum in reality, caused by 
sample inhomogeneities and by truncation errors in the numerical 
evaluation of $\sigma(T = {\rm const.},n)$ utilizing a difference 
quotient. 

The hypothesis of the sharp maximum in 
$\mbox{d}\;\!\mbox{log}_{10}\;\!\sigma / \mbox{d} n$ is checked here 
using the same approach as in Eqs.\ (3), (8), and (9) of Ref.\ 
\onlinecite{Moe.etal.99}: A window including 4 neighboring points is 
shifted along  the $\mbox{log}_{10}\;\!\sigma$ curve. The slope values 
from the corresponding linear fits of $\mbox{log}_{10}\;\!\sigma$ as 
function of $n$ are related to the $n$ values, for which polynomials of 
second order would be exactly differentiated. To consider 4 points in 
the fits turned out to be a good compromise between the demands for high 
resolution and for low random errors.

Fig.\ 9 shows that, independent of $T$,
$\mbox{d}\;\!\mbox{log}_{10}\;\!\sigma / \mbox{d} n$ exhibits the 
expected sharp maximum -- note that a logarithmic scale is used to 
display the derivative values --. It is striking that the peak is always 
located just at $n_{\rm c}$. This feature is another indication of phase 
transitions occurring at finite $T$ when $n$ crosses $n_{\rm c}$. 

From the phenomenological point of view, the peak of the derivative has 
two effects: On the one hand, it is the reason of the small differences 
between $n_{\rm c}$ and the $n_{\rm k}$ values from the piecewise linear 
fits, see Sec.\ II. On the other hand, it makes the sharp bend in 
$\sigma(T = {\rm const.},n)$ more easily visible.

The question remains, whether the $T$ dependence of the amplitude of the 
rounded step can also be understood. As discussed above, the amplitude 
of the ``rounded step'' decreases with increasing $T$ as expected from 
Eqs.\ (6) and (7). However, for a quantitative analysis, more detailed 
experimental data are needed.
 
Finally, a counterintuitive consequence of the scaling is worth pointing 
to: In measuring $\sigma(t)$, the influence of sample inhomogeneities 
decreases with increasing $T$ since the $n$ range needed to explore a 
certain fixed $\sigma$ interval broadens. Hence measurements at 
intermediate $T$ should be more promising than studies at extremely low 
$T$.

\section{Conclusions}

In summary, reconsidering the experiment of Lai and 
coworkers,\cite{Lai.etal.07} a series of features has been described
which indicate that the apparent MIT at $T = 0$ may be connected with a 
line of sharp continuous phase transitions at finite $T$: (a) Close to 
the critical charge carrier concentration $n_{\rm c}$, defined by the 
sign change of $\mbox{d} \sigma / \mbox{d} T$ as $T \rightarrow 0$, 
piecewise linear functions approximate $\sigma(T = {\rm const.},n)$ 
clearly better than polynomials of third order. This holds for the 
entire $T$ range from 0.35 to $4.0\ {\rm K}$. (b) The knee of the 
piecewise linear function is always located close to $n_{\rm c}$. (c) 
In the immediate vicinity of $n_{\rm c}$, $\rho(T = {\rm const.},n)$ has 
a ``rounded step'' structure of $T$-independent width what is connected
with an offset in the curves. (d) The middle of the ``rounded step'' 
always coincides with $n_{\rm c}$. (e) Two previous 
experiments\cite{Krav.etal.95,Papa.Shay} exhibit similar offsets in 
$\rho(T = {\rm const.},n)$ close to $n_{\rm c}$. (f) The offsets in 
the data from Refs.\ \onlinecite{Lai.etal.07,Krav.etal.95}, and 
\onlinecite{Papa.Shay} occur at roughly the same $\rho$ value, a little 
below $h/e^2$. (g) The data from Ref.\ \onlinecite{Lai.etal.07} exhibit 
sharp peaks in the derivative 
$\mbox{d}\;\!\mbox{log}_{10}\;\!\sigma / \mbox{d} n$ at $n_{\rm c}$ for 
all considered $T$ values. 

Moreover, it has been shown that the ``rounded step'' in the
$\rho(T = {\rm const.},n)$ curves can be explained in terms of scaling 
of the $T$ dependences of $\sigma$ for various $n$ within the region of 
activated conduction. The applicability of this scaling is supported by 
the collapse of the curves  for $T = 0.35$ and $0.64\ {\rm K}$ when 
plotting $\sigma$ as function of $(n - n_{\rm c})/T^{0.625}$, utilizing
the value of the critical exponent of the characteristic temperature 
obtained in Ref.\ \onlinecite{Krav.etal.95}. Additionally, an 
approximate collapse is observed for the whole $T$ region from 0.35 to 
$4.0\ {\rm K}$ when considering the ratio 
$\sigma(T,n) / \sigma(T,n_{\rm c})$ rather than $\sigma(T,n)$. 

The scaling of $\sigma(T,n)$ for the region of activated conduction 
should cease to be valid when the characteristic hopping temperature 
reaches zero indicating a qualitative change in the transport mechanism. 
This happens at the $T$-independent resistivity value, 
$\sigma_{\rm c} = \sigma(T,n_{\rm c})$. In this sense, scaling gives 
additional, indirect support to the hypothesis of a line of finite 
temperature phase transitions from activated to apparent metallic 
conduction. 

As a whole, the above indications seem to provide rather convincing 
evidence for a line of sharp phase transitions at finite $T$. 
Nevertheless, to the best of my knowledge, there is no microscopic 
theory available explaining this feature. So further theoretical work as 
well as additional experiments in the immediate vicinity of the apparent 
MIT are called for. These experiments should primarily focus on 
enhancing precision, in particular with respect to homogeneity of the 
samples, rather than on reducing the lowest accessible temperature. If 
they turn out to support the above conclusion, they will shed new light 
on the problem whether or not the apparent MIT is a real phase 
transition, and set qualitative constraints on theoretical models. In 
such investigations, scaling analyses as performed above may be helpful 
to clarify the character of the apparent metallic state.

\begin{acknowledgments}
The critical comments by C.J.~Adkins on Ref.\ \onlinecite{Moe.02} and 
on this work were very helpful in formulating the publication. In 
particular, I am indebted to him for motivating the high temperature
part of the scaling analysis in Figs.\ 7 and 8. I am much obliged to
M.~Schreiber for numerous remarks on the manuscript, particularly for 
demanding a neat treatment of the problem of the unknown knee position 
in the piecewise linear fits. Moreover, many helpful discussions with 
K.~Morawetz, M.~Richter, and W.~Schirmacher are gratefully acknowledged.
\end{acknowledgments}

\end{document}